\documentclass[%
 aip,
 jmp,%
 amsmath,amssymb,
 reprint,%
]{revtex4-2}

\usepackage{comment}
\usepackage{graphicx}
\usepackage{dcolumn}
\usepackage{bm}
\usepackage{times}
\usepackage{helvet}

\begin{document}

\title[Filters reveal emergent structure in computational morphogenesis]{Filters reveal emergent structure in \\computational morphogenesis}

\author{Hazhir Aliahmadi}
 
\author{Aidan Sheedy}

\author{Greg van Anders}
\email{gva@queensu.ca}
\affiliation{ 
Department of Physics\text{,} Engineering Physics and Astronomy, Queen's University\text{,} Kingston ON\text{,} Canada
}%

\date{\today}

\begin{abstract}
Revolutionary advances in both manufacturing and computational morphogenesis
  raise critical questions about design sensitivity. Sensitivity questions are
  especially critical in contexts, such as topology optimization, that yield
  structures with emergent morphology. However, analyzing emergent structures via
  conventional, perturbative techniques can mask larger-scale vulnerabilities
  that could manifest in essential components. Risks that fail to appear in
  perturbative sensitivity analyses will only continue to proliferate as topology
  optimization-driven manufacturing penetrates more deeply into engineering
  design and consumer products. Here, we introduce Laplace-transform based
  computational filters that supplement computational morphogenesis with a set
  of nonperturbative sensitivity analyses. We demonstrate how this approach
  identifies important elements of a structure even in the absence of knowledge
  of the ultimate, optimal structure itself. We leverage techniques from
  molecular dynamics and implement these methods in open-source codes,
  demonstrating their application to compliance minimization problems in both 2D
  and 3D. Our implementation extends straightforwardly to topology optimization
  for other problems and benefits from the strong scaling properties observed in
  conventional molecular simulation.
\end{abstract}

\maketitle

\section{\label{Sec:Introduction}Introduction}
Computational morphogenesis is a cornerstone of engineering practice, most
notably as enacted via iterative methods for optimizing morphology such as
topology optimization \cite{bendsoeGeneratingOptimalTopologies1988, TopOptBook}.
Recently, topology optimization algorithms have been propelled by major
computational advances to unprecedented resolution
\cite{gigavoxel,alexandersenLargeScaleThreedimensional2016,aageTopologyOptimizationUsing2015}.
The rise of ultra-high resolution topology optimization algorithms raises
important questions about the comparative performance of as-designed optimal
solutions and as-realized physical components: if design precision outstrips
manufacturing precision, what forms of deviation between design and realization
are permissible (or not) from a performance
perspective?\cite{lazarovLengthScaleManufacturability2016, TopOptAddMfg}
Arguably, precision advances bring pre-existing design/manufacturing contrast
questions into sharper relief. Manufacturing processes typically involve
trade-offs between fidelity and throughput, meaning that no design is ever
realized with absolute precision in practice.

The ubiquitous discrepancy between design and realization begs for improved
understanding, but is complicated by the combination of optimization and
emergence. Optimization algorithms are effective at generating solutions, but
are less appropriate tools for anticipating real-world performance. Attempts
to predict the behaviour of realizations sometimes simulate the physical
properties of optimized designs. Using optimal designs to forecast the behaviour
of non-optimal realizations is problematic, however, because the analysis is
predicated on the anticipation of discrepancies in key features between the
hypothetical design and physical reality. Discrepancies between designs and
realizations limit the validity of extrapolating from optimal designs to
non-optimal designs that are realized in practice. The critical role of
non-optimal solutions in anticipating the behaviour of realizations demands new
approaches that centre the analysis of non-optimal designs.

New approaches to that address design--realization discrepancies are even more
critical because of the conundrum raised by structures with emergent morphology:
Computational morphogenesis approaches such as topology optimization provide the
greatest design value when they yield structures with emergent morphology.
Designs with emergent morphology epitomize the value of topology optimization
because the algorithm yields solutions even when the problem formulation
provides few clues to the ultimate solution. However, the obscure connection
between ultimate, emergent morphology and the original design problem impedes
intuitively identifying critical components.

To understand how to make systematic design--realization connections, it is
crucial to identify which components of the expected solution contribute most to
the design's purpose. Yet, purpose-based component identification cannot solely
rely on perturbative analyses. Perturbative approaches to sensitivity
\cite{choDesignSensitivityAnalysis2003,castilloSensitivityAnalysisOptimization2008}
and trade-off \cite{hegwoodWhyWinWins2022} analyses tend to be limited by
the searching method and initial conditions. These limitations indicate the need
for more comprehensive, non-perturbative approaches. One set of non-perturbative
approaches deployed in other domains of engineering is provided by integral
transforms that offer a global perspective on the function
being analyzed \cite{debnathIntegralTransformsTheir2007}.
For example, the frequency domain techniques \cite{zadehTheoryFiltering1953}
that pervade control engineering \cite{ogataModernControlEngineering2010} and
signal processing \cite{oppenheimSignalsSystems1983} paint a comprehensive
picture of underlying behavioural drivers. The success of integral transform
techniques in these problems raises the question of whether adaptations exist
for non-perturbative analyses of emergent morphogenesis.

Here, we show that Pareto-Laplace transforms \cite{ParetoLaplace} of the
solution space geometry of a typical computational morphogenesis problem,
compliance minimization \cite{TopOptBook}, yield crucial insights into
design--realization relationships and the emergence of structure in
computational morphogenesis. Analogous to the time-frequency duality present in
problems in signal processing \cite{oppenheimSignalsSystems1983}, we find that
key morphological features are structural analogs of long-lived transients. We
make a precise mathematical identification of the Laplace transform of the solution
space volume as a partition function in statistical physics \cite{ParetoLaplace}.
We leverage this identification to implement the transform using molecular
dynamics frameworks \cite{frenkelsmit} that are industry-standard practice in
chemistry, physics, and materials science.

We use a statistical physics mapping to show that the geometry of the
solution space encodes the emergence of morphology in a discrete set of regimes
encoded by temperature in the transformed space. We show that this
identification both rationalizes aspects of the emergent structures through a
series of condensation regimes and provides a mapping of regions of the system
by their importance for overall structural performance. An advantage of our
filter-based approach to computational morphogenesis is that it makes it
possible to identify the critical elements of an ultimate design without the
need to know the full, optimized design itself.

For concreteness, we show results for compliance minimization that we use to
compare directly with widely-known, open-source implementations of topology
optimization in 2D \cite{andreassenEfficientTopologyOptimization2011} and 3D \cite{aageTopologyOptimizationUsing2015}. We provide
open-source implementations of our methods in MATLAB \cite{Hyp88} and C++/Python
\cite{HypOptLib}. Although we give specific results of our methods applied to
problems in compliance minimization, our method generalizes straightforwardly to
other problems in topology optimization where the design objectives rely on, e.g.,
electromagnetic, optical, chemical, thermodynamic, hydrodynamic, or acoustic
phenomena. 
\section{\label{Sec:Results}Results}
\subsection{\label{SubSec:MorphologyEmergesViaDistinctRegimes}Morphology Emerges Via Distinct Regimes of Structural Condensation}
\begin{figure}[!h]
\centering
  \includegraphics[width=1.0\textwidth]{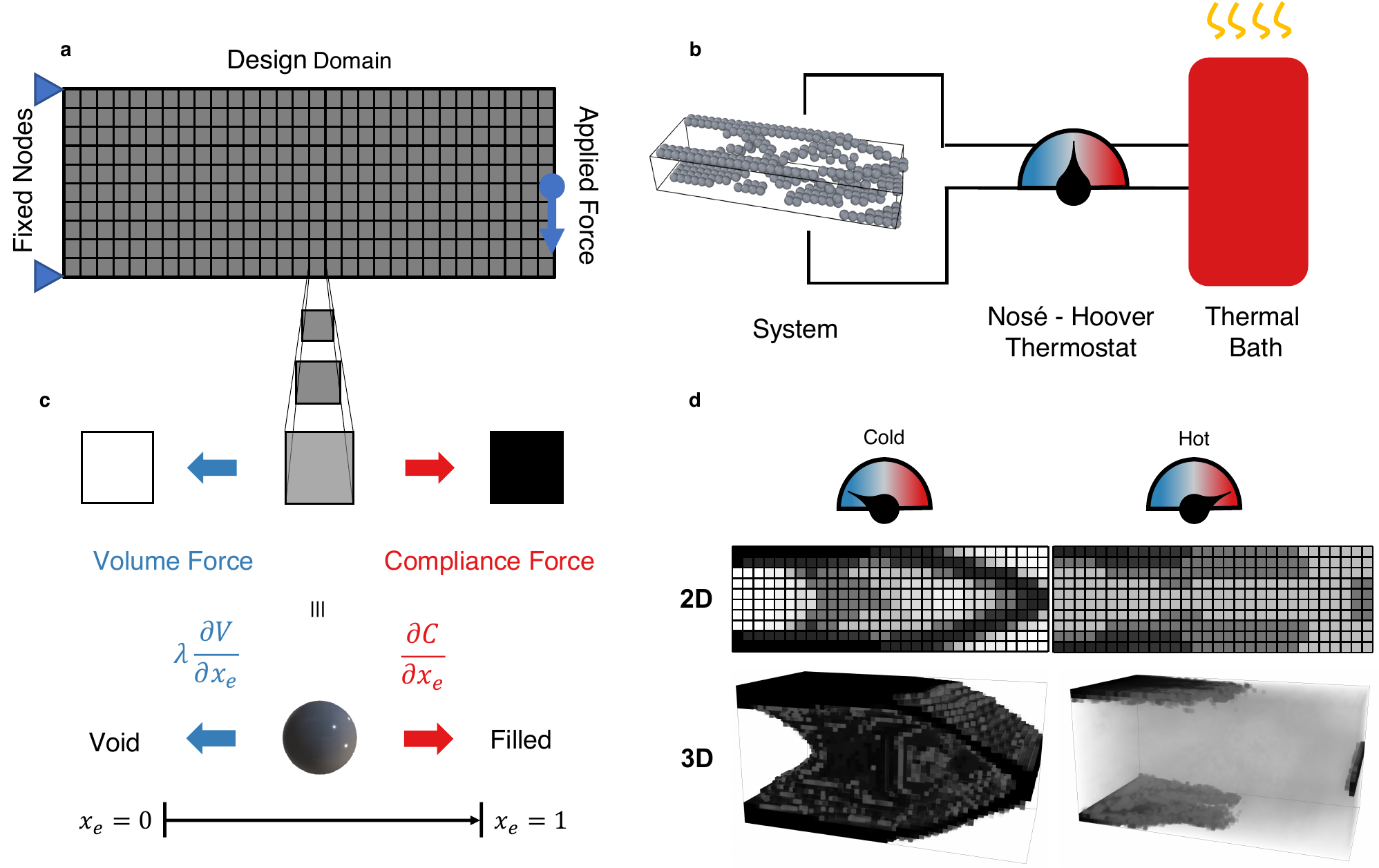}
  \caption{
    \textbf{The Pareto-Laplace transform for compliance minimization:
    density dynamics and ensemble generation.} (\textbf{a}) Compliance
    minimization is about distributing a limited amount of material within a
    Design Domain to minimize compliance (deformation) under Applied Forces,
    considering boundary conditions like Fixed Nodes. The density of each site,
    $x_e$, is a value between $0$ and $1$, where a higher density reduces
    compliance, while a lower density reduces total used material (volume).
    (\textbf{c}) The density of each site is analogized to the position of a
    particle, where forces from compliance and volume
    constraints influence its position. (\textbf{b}) Connecting the particle
    system to a heat bath via a Nos\'e-Hoover thermostat generates an
    isothermal ensemble of design solutions. (\textbf{d}) shows the average design solutions for low and
    high-temperature ensembles in 2D and 3D cantilever beam problems. 
  }
  \label{Fig:ConceptFigure}
\end{figure}

Fig.\ \ref{Fig:ConceptFigure} illustrates the Pareto-Laplace transform for a topology optimization problem, such
as compliance minimization for cantilever beam design, using molecular dynamics. We formulate a
``molecular'' version of the problem by writing a Hamiltonian representing each design domain site as a particle. We assign particles a
``position'' between 0 (void of material) and 1 (filled with material). The
compliance objective generates a gradient force that imparts kinetic energy to
particles, which we control using a Nos\'e-Hoover thermostat. We sample the landscape at a fixed temperature to generate
collections of solutions (i.e., a canonical ensemble of solutions
\cite{frenkelsmit}), and these solutions elucidate the structure of the
solution landscape. (Refer to SI Movie [1] for an animation illustrating a molecular approach to compliance minimization in cantilever beam design at a low temperature.)

\begin{figure}[!h]
\centering
  \includegraphics[width=1.0\textwidth]{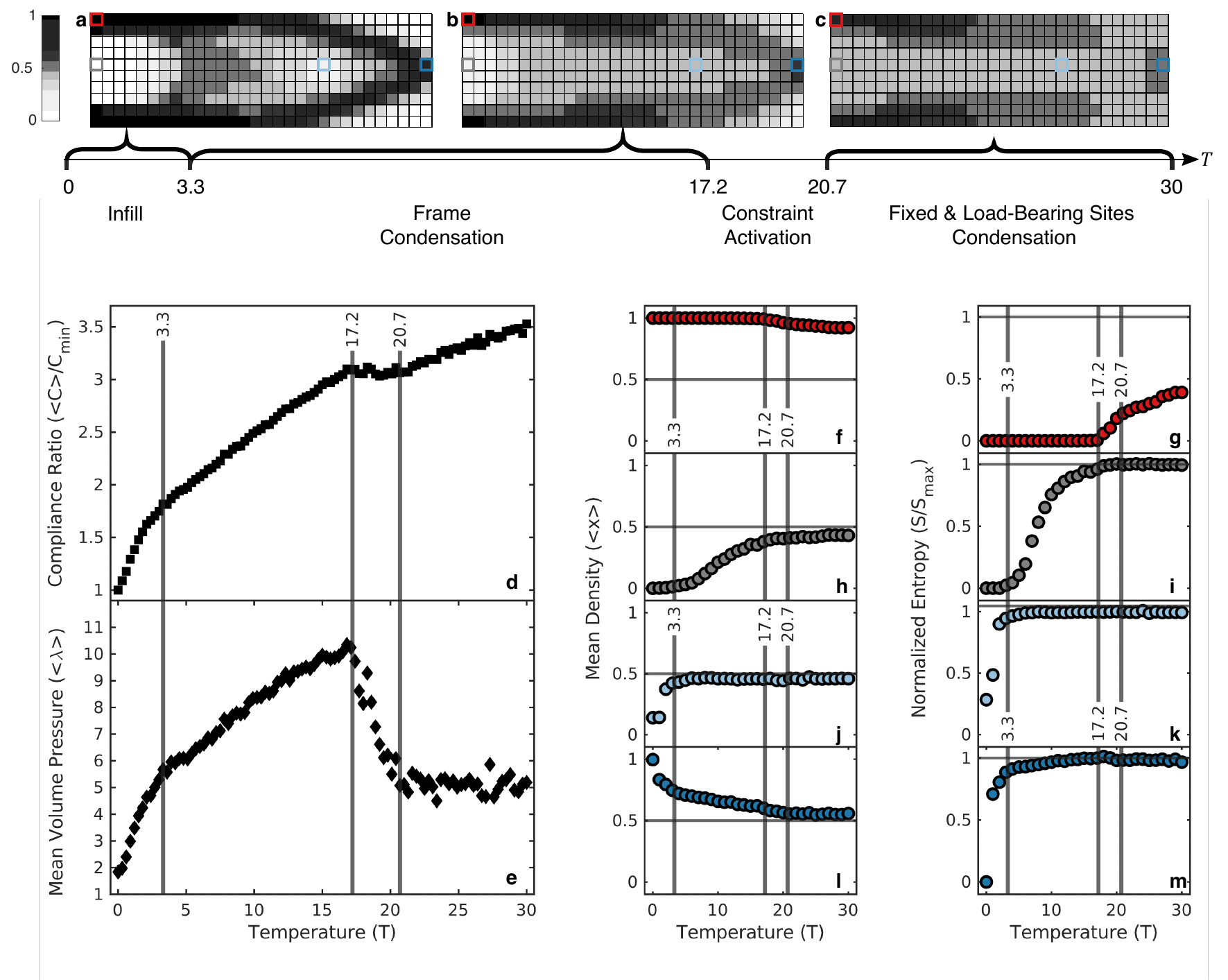}
  \caption{
    \textbf{Three Phases of Condensation for a 2D Cantilever Beam Problem.} 
    \textbf{(a–c)} mean material distribution at representative temperature regimes: 
    \textbf{(a)} $T \lesssim 3.3$, showing infill condensation within the frame; 
    \textbf{(b)} $3.3 \lesssim T \lesssim 17.2$, illustrating the formation of a connecting frame;
    \textbf{(c)} $T > 20.7$, highlighting initial condensation near fixed and load-bearing nodes. 
    \textbf{(d–e)} Compliance ratio $\langle C \rangle / C_\text{min}$ and mean pressure $\langle \lambda \rangle$ as functions of temperature, respectively.
    For $17.2 \lesssim T \lesssim 20.7$, the mean volume pressure increases sharply, while the compliance ratio remains nearly constant, indicating that the reduction in compliance is due to the addition of material rather than condensation.
    \textbf{(f–m)} Site-specific Mean density $\langle x \rangle$ and entropy density $S/S_\text{max}$ across four representative sites: \textbf{(f, g)} An essential site to compliance minimization (red square in \textbf{(a–c)}), exhibiting high density and low entropy even at high $T$. \textbf{(h, i)} A site with a contribution to volume reduction rather than compliance minimization (gray square in \textbf{(a–c)}), characterized by simultaneous loss of density and entropy at $T \approx 20$. \textbf{(j, k)} A designable site (light blue square in \textbf{(a–c)}) maintains high entropy and $\langle x \rangle$ until $T \approx 3.3$. \textbf{(l, m)} A sensitive site (dark blue square in \textbf{(a–c)}) converges to its final density at higher $T \approx 20$ while preserving entropy down to $T \approx 3.3$.}
  \label{Fig:CondensationPhases}
\end{figure}

From collections of solutions, Fig.\ \ref{Fig:CondensationPhases} shows that
low-compliance 2D structures emerge in a process that is analogous to
condensation, which occurs via three distinct regimes demarcated by temperature
(see Fig.\ \ref{Fig:Phases3D} for the 3D cantilever beam). We identified these
regimes by computing the compliance ratio, defined as the ratio of the mean
compliance of sampled solutions over the solution ensemble to the minimum
compliance, \( \langle C \rangle / C_\text{min} \), and the mean pressure \(
\langle \lambda \rangle \) required to maintain a total material (volume)
fraction \( V_f = 0.5 \). These metrics are presented as functions of
temperature in Fig.\ \ref{Fig:CondensationPhases}d and Fig.\
\ref{Fig:CondensationPhases}e, respectively.

For the regime $T\gtrsim 20.7$ (all temperatures are given in units of
compliance) the first structural features to condense are adjacent sites to the
fixed and load-bearing nodes in the cantilever beam problem (see Fig.\
\ref{Fig:CondensationPhases}c). We find that this regime coincides with
solutions that exhibit $\langle C \rangle/C_\text{min}\gtrsim 3$, i.e., three
times the minimum possible value for the
compliance. 

For $T\approx20.7$ we observe a change in the slope of both $\langle C \rangle/C_\text{min}$ and
$\lambda$ that signals the onset of constraint activation. Between
$17.2\lesssim T\lesssim 20.7$ the material deposited near the fixed sites
remains constant. We note that this behaviour coincides with a steep
increase in the mean volume pressure $\langle \lambda\rangle$ (Fig.\ \ref{Fig:CondensationPhases}e) and a near-zero slope for
$\langle C \rangle/C_\text{min}$ (which remains at $\langle C \rangle/C_\text{min}\approx 3$, Fig.\ \ref{Fig:CondensationPhases}d). These
results indicate that negligible morphological change occurs because the
system is increasingly driven to reduce compliance by adding material rather
than by condensing existing material.

We find that solutions with $2\lesssim \langle C \rangle/C_\text{min}\lesssim 3$ emerge for
$3.3\lesssim T\lesssim 17.2$, and this reduction in compliance coincides
with the condensation of a ``frame'' connecting the fixed- and load-bearing sites.
For $T\lesssim 3.3$, we find a near halving of the compliance as it converges to
$\langle C \rangle/C_\text{min}=1$, which coincides with the condensation of ``infill'' inside
the frame.

\subsection{\label{SubSec:OptimalSolutionsDifferentRoles}Optimal Solutions Depend on All Sites but with Different Roles}
The emergence of morphology via a series of distinct regimes of structural
condensation can be ascribed to a corresponding set of microscopic,
site-specific behaviours. Fig.\ \ref{Fig:CondensationPhases} panels f,h,j,l show mean material density (referred to as mean density hereafter), $\langle x\rangle$, as a
function of temperature for four representative sites (corresponding data for
all sites are given in Fig.\ \ref{Fig:DensityDynamics}) and Fig.\ \ref{Fig:CondensationPhases} panels g,i,k,m show entropy density, $S/S_{max}$,
as a function of temperature for the same four representative sites
(corresponding data for all sites are given in Fig.\ \ref{Fig:EntropyDynamics}). This analysis reveals a range of sensitivity among sites in minimizing compliance, where some sites rely on the contributions of other sites and the overall design, while others act independently and are essential for achieving compliance reduction. (Refer to Appendix \ref{AppSec:EntropyCalculation} for details on entropy calculation.)

Fig.\ \ref{Fig:CondensationPhases}f,g show the mean density and entropy density of a representative site that is essential to minimizing compliance. This site (identified by a red square in Fig.\ \ref{Fig:CondensationPhases}a,b,c) has high
mean density, $\left<x\right>\approx 1$, and low entropy density,
$S/S_\text{max}<0.5$, even at high temperatures, $T\gtrsim 20$. However, low entropy density at high temperatures is insufficient to conclude a site's role in the overall design. Fig.\ \ref{Fig:CondensationPhases}h,i correspond to
a representative site (indicated by a grey box in Fig.\ \ref{Fig:CondensationPhases}a,b,c) that makes
a negligible contribution to overall compliance minimization. This is signaled by a coincident loss in entropy density and in mean density
beginning at relatively high temperature $T\approx 20$. In effect, these sites
contribute to the design by reducing volume rather than compliance.

In contrast to the sites marked with red and gray squares in Fig.\ \ref{Fig:CondensationPhases}a,b,c, some sites reach maximum entropy at much lower temperatures. The behaviour of these sites can be classified by considering their mean density. The sites, like the one marked with a light blue square in Fig.\ \ref{Fig:CondensationPhases}a,b,c, are designable, which maintain high entropy density $S/S_{\text{max}} \approx 1$, and mean density $\langle x \rangle \approx 0.5$ down to relatively low temperatures, $T \approx 3.3$ as shown in Fig.\ \ref{Fig:CondensationPhases}j,k. The other ones, like the site indicated by a dark blue square in Fig.\ \ref{Fig:CondensationPhases} a,b,c, are sensitive and start to converge to their final density at high temperature ($T \approx 20$ in this case), yet preserve high entropy down to low temperature ($T \approx 3.3 $ in this case).

\subsection{\label{SubSec:CondensationTemperatureMap}Condensation Temperature Map Signals Site-Specific Importance}
Site-specific results reported in Fig.\ \ref{Fig:CondensationPhases} indicate that although all sites
contribute to minimizing compliance at a fixed volume, different sites play very different roles. We associated roles with quantifiable behaviours, and we can further
assign relative importance to individual sites.

Fig.\ \ref{Fig:CondensationPhases} provided a set of indicators that signal distinct roles for different
sites. However, reporting these indicators for all sites would be cumbersome,
particularly for extensions to compliance minimization in three dimensions, so
it is instructive to use a proxy. To define this proxy, we note Fig.\ \ref{Fig:CondensationPhases}
indicates a clear association between overall compliance and temperature, so we
classify importance using condensation temperature for each site. We defined the
condensation temperature for a site as the temperature at which $S/S_\text{max}<0.85$, where the site's density behaves like a gas.

\begin{figure}[!h]
\centering
  \includegraphics[width=1.0\textwidth]{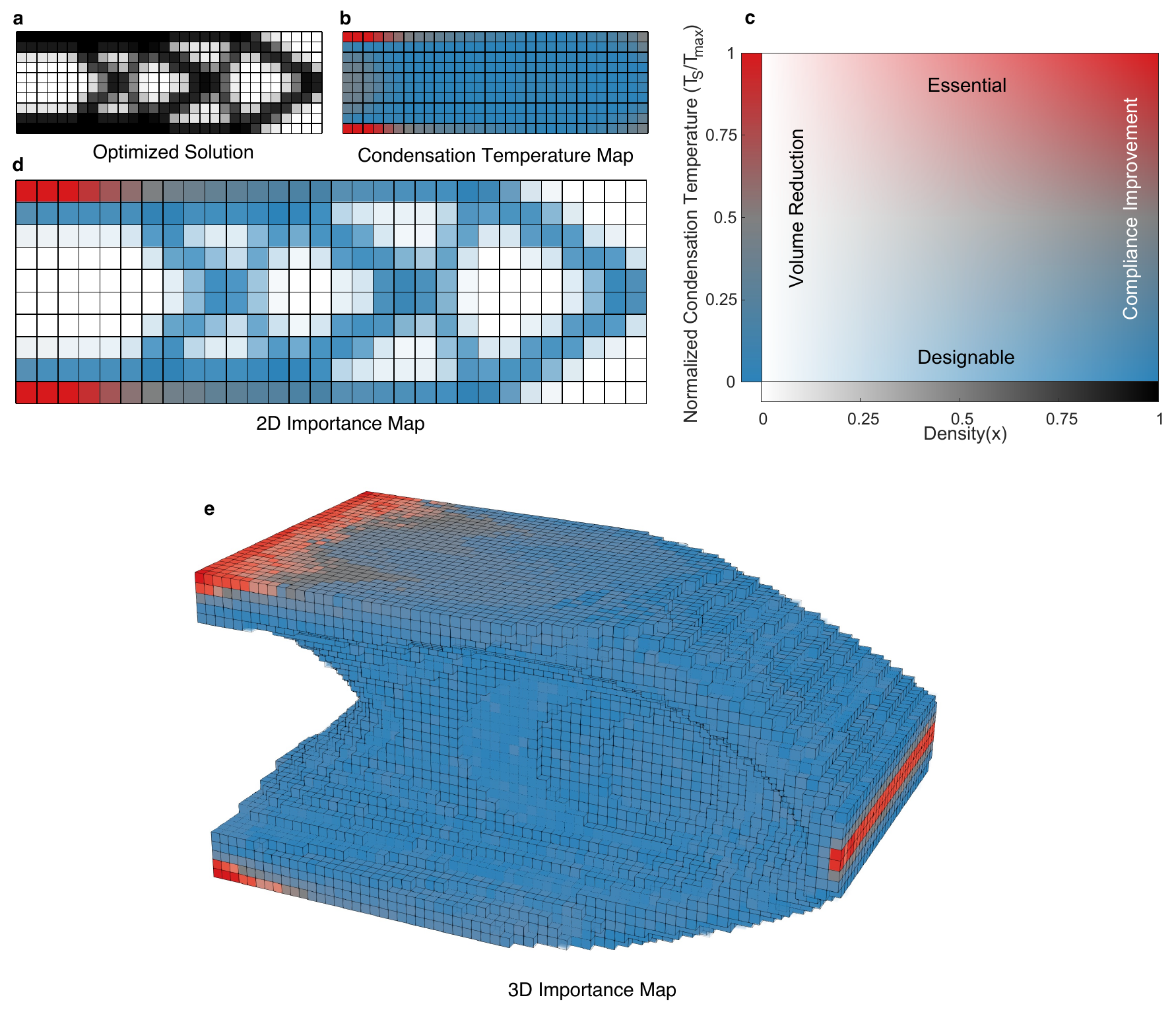}
  \caption{
  \textbf{Importance Maps for 2D and 3D Cantilever Beam Design
  Problems.}(\textbf{a}) An optimized solution as solved by
  \cite{andreassenEfficientTopologyOptimization2011}. (\textbf{b}) The
  Condensation Temperature Map, showing the normalized condensation temperature
  across the design domain with the colormap in (\textbf{c}), assuming a filled
  density ($x_e=1$) for all sites. (\textbf{c}) shows a 2D colormap where the
  horizontal axis represents site density and the vertical axis shows the
  normalized condensation temperature. The vertical axis indicates site
  importance, with higher condensation temperatures marking essential sites and
  lower temperatures indicating designable ones. High-density sites (opaque)
  contribute to compliance minimization, while low-density sites (transparent)
  aid in volume reduction.(\textbf{d}) The Importance Map for the 2D Cantilever
  Beam, based on the optimized solution in (\textbf{a}) and the Condensation
  Temperature Map in (\textbf{b}), using the colormap provided in (\textbf{c}).
  (\textbf{e}) The Importance Map for the 3D Cantilever Beam.
  }
  \label{Fig:ImportanceMap}
\end{figure}

Fig.\ \ref{Fig:ImportanceMap} shows importance maps for compliance minimization of a cantilever beam
in two- and three dimensions. For the 2D problem, Fig.\ \ref{Fig:ImportanceMap}a shows an optimized solution adjacent to the condensation
temperature map, see Fig.\ \ref{Fig:ImportanceMap}b. Fig.\ \ref{Fig:ImportanceMap}d depicts the importance map that results
from composing the optimal solution and the condensation temperature map using
the colour density map depicted in Fig.\ \ref{Fig:ImportanceMap}c. The most important sites, indicated
in red, are sites that condense at very high temperatures, i.e., those sites are essential to be filled even for designs with a moderate level of compliance. Less important sites, shown in blue, condense only at low temperatures. These sites are considered designable; although they are included in the optimized solution, manufacturing errors affecting their density do not compromise the optimality of the design as much as for red or gray sites.

Fig.\ \ref{Fig:ImportanceMap}e gives a corresponding importance map for a 3D cantilever beam. Like the
2D case, results indicate that the sites adjacent to the fixed sites are among
the most important. However, in the 3D case, this importance is shared by the
load-bearing sites. (Refer to SI Movie [2] for an animation showcasing the 3D map from all angles, and see Fig. \ref{Fig:ImportanceMap3D} for further details on the importance map related to the 3D cantilever beam problem.)

\subsection{\label{SubSec:EffectiveDimensionality}Effective Dimensionality Facilitates Overall Design Characterization}
\label{sec:EffDim}
We observed that structures condense non-uniformly as a function of
temperature to reduce compliance in a series of regimes. Each regime that
results in a compliance reduction is associated with a set of sites. We mapped how
compliance reduction depends on different sites via the importance maps
shown in Fig.\ \ref{Fig:ImportanceMap}. However, it is easy to imagine cases where it is important to extract broader-level design information in
addition to the granular understanding of a single design. The behaviour of the compliance--temperature response signals the underlying geometry of the solution space which is a key indicator for this class of comparisons.

\begin{figure}[!h]
\centering
  \includegraphics[width=1.0\textwidth]{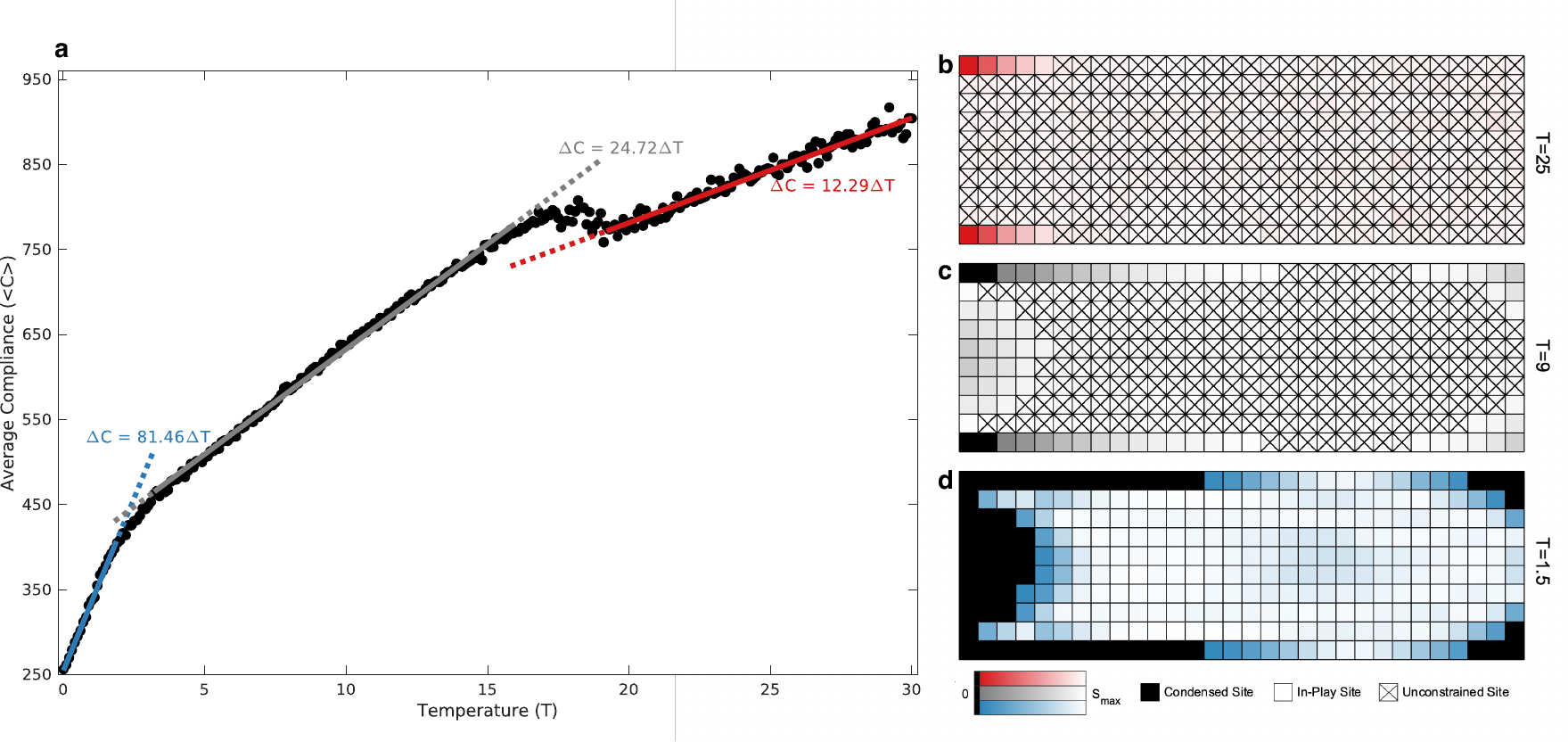}
  \caption{
    \textbf{Linear Response of Average Compliance to Changing Temperature.}
    (\textbf{a}) Different slopes of the linear response indicate distinct
    regimes of condensation, each with a different number of effective dimensions.
    (\textbf{b}) At $T=25$, most sites are unconstrained with maximized entropy, and only
    the few in-play sites  contribute to reducing compliance, resulting in the
    linear behavior is shown by the red line in (\textbf{a}). As temperature
    decreases to $T=9$, more sites become condensed, leading to more in-play sites and a higher slope of the linear response, as illustrated by
    (\textbf{c}) and the grey line in (\textbf{a}). (\textbf{d}) shows the
    design domain at $T=5$, where most sites are in-play, optimizing the infill
    design to minimize compliance.
  }
  \label{Fig:LinearResp}
\end{figure}

Fig.\ \ref{Fig:LinearResp} gives an annotated plot of $\left<C\right>$ vs $T$ for a 2D
cantilever beam. As we noted above, the condensation regimes are indicated by a
series of approximately linear responses. In conventional physical
systems, $\left<C\right>(T)$ would be referred to as an equation of state, and
its form would reflect the structure of the underlying state space. Here, the
state space is the solution space, and so the behaviour of the solution space must
signal how optimized designs emerge.

In Methods (see, Sec.\ \ref{SubSec:NOE}) we use the Pareto-Laplace transform to
derive a theory that relates the slope of the compliance--temperature to the growth of the number of candidate designs with fixed,
non-minimal compliance. We consider a regime where the number of candidate designs, denoted as 
\(\Omega\), follows the scaling relation
\begin{equation}
    \Omega(C) \sim (C-C_{min})^{\frac{N_{IP}}{\nu} - 1} \;,
\end{equation}
where \(\Omega(C)\) increases with \(N_{IP}\), the effective number 
of ``in-play'' dimensions. Here, \(N_{IP}\) represents the number of sites 
capable of participating in a correlated exchange of material while maintaining 
the fixed level of mean compliance under a constant total volume constraint. The parameter 
\(\nu\) serves as a growth index; for instance, if the volume of solution space scales linearly 
with the number of sites, \(\nu = 1\). In Section~\ref{SubSec:BNOD}, we demonstrate that this growth leads to the 
following expression for the compliance--temperature slope:
\begin{equation}
    \frac{\partial \langle C \rangle}{\partial T} = \frac{N_{IP}}{\nu} \; ,
    \label{eq:Cslope}
\end{equation}
where this slope encodes the effective number of ``in-play'' sites 
at a given compliance level.

We derive Eq.\ \eqref{eq:Cslope} in Sec.\ \ref{SubSec:NOE}, and \ref{SubSec:BNOD}, but to see how this
arises on more than purely mathematical grounds, it is useful to form an
approximate grouping of sites into one of three, temperature-dependent states:
unconstrained, in-play, or condensed. At a given temperature, condensed sites, akin to particles in a
solid, have essentially fixed material
density and vanishing entropy density. Unconstrained sites, akin to particles in a gas, have
$\left<x\right>\approx 0.5$ and maximal entropy density, and fluid-like in-play sites are in the process of condensing and constitute degrees of
freedom that affects the solution space at that temperature.

Condensed
sites do not contribute to $\Omega(C)$, and therefore do not
contribute to $\left<C\right>(T)$, because their degrees of freedom are
effectively ``frozen out''. Unconstrained sites also contribute to neither
$\Omega(C)$ nor $\left<C\right>(T)$, however rather than being frozen out,
those gas-like sites are unaffected by the objective function or constraints at
a given $T$. The only sites that do contribute are the fluid-like,
in-play sites, that engage in a correlated exchange of material to maintain some
fixed level of mean compliance for a constant total volume.

The slope of $\left<C\right>$ vs $T$ in \ref{Fig:LinearResp}a provides a direct
estimate of the relative size of regions involved in the formation of structure
at a fixed level of compliance. For the 2D cantilever beam, we are able to
confirm this site classification directly by mapping individual sites in Fig.\
\ref{Fig:ImportanceMap}b,c,d (Refer to SI Movie [3] for an animation illustrating how the sites condense as the temperature decreases.). However, this approach could also be repeated for larger-scale
systems with large numbers of sites where the granular
visualization of individual sites is impractical.

\section{\label{Sec:Discussion}Discussion}
We showed that integral transforms of solution spaces in topology optimization
provide new forms of insight into design--realization contrasts in computational
morphogenesis. Our approach supplements powerful, widely-used methods for
topology optimization, which emerge as a zero-temperature limit of our approach.
Though we give explicit results for problems in compliance minimization at
scales that facilitate comparison with well-known results in topology
optimization, our methods are straightforward to extend.

Topology optimization has been demonstrated for a wide range of problems, e.g.,
material
design\cite{andreassenDesignManufacturable3D2014,gaoTopologicalShapeOptimization2018,xuIsogeometricTopologyOptimization2020},
photonic structures,\cite{TopOptPhotonic, frellsenTopologyOptimizedMode2016},
microfluidics,\cite{zhouVariationalLevelSet2008, dengTopologyOptimizationUnsteady2011,yoonNewMonolithicDesign2023} heat transfer,\cite{alexandersenLargeScaleThreedimensional2016, alexandersenTopologyOptimisationNatural2014}, among many others. The key
feature that makes topology optimization portable across all of these domains is
that in each application, the only difference is the underlying
partial differential equation that links solution morphology to the design
objective. Our Nos\'e-Hoover thermostat-based, molecular dynamics implementation
of the compliance minimization Laplace transform is, in essence, a ``superset''
of gradient-based topology optimization. Any gradient-based topology
optimization problem could be ``lifted'' using the same set of mathematical
procedures we employ here. Indeed, we are engaged in molecular dynamics-based
extensions of the present methodology to topology optimization problems in other
domains, which we hope to report on soon.

Although the specific numerical results we report here were computed with a
molecular dynamics-based algorithm, the insights they provide stem from the more
foundational structure which is the Pareto-Laplace transform
\cite{ParetoLaplace} of the solution space geometry. Laplace transforms of the
sort we derive here have been employed in statistical physics for more than a
century, and advanced methods have been derived for computing them in cases
where there is no well-defined notion of a gradient. Consequently, we anticipate
that the proposed approach will extend to non-gradient topology optimization
problems.

We gave examples of filter-based analyses using example codes in MATLAB
\cite{Hyp88} and a PetSc-based C++ implementation \cite{HypOptLib}. However,
because the approach relies on molecular dynamics techniques, other
implementations of filter-based analyses could deploy advanced sampling methods
\cite{frenkelsmit} or could take advantage of the strong scaling of molecular
dynamics to thousands of graphics processing units \cite{hoomdmpi}. We should
also note that although we used a Nos\'e-Hoover thermostat, other methods of
thermostatting, e.g., via a Langevin thermostat
\cite{davidchackLangevinThermostatRigid2009}, might find use in other settings.

Finally, we note that the filter-based approach we present above show it is
possible to predict a design's most important elements without knowing the final
design's full form. This signals that filters could be deployed as a leading
indicator in an iterative, complex design process. In addition, although we
focused on the emergence of structure through condensation, applying our filter
approach to ``evaporate'' designs could provide important training data for
generative AI methods for engineering structures that mimic techniques used for
image generation.

\section{\label{Sec:Methods}Methods}
\subsection{\label{SubSec:ComplianceMin}Compliance Minimization}
Topology optimization problems typically involve the deposition of material in a
fixed design domain according to some optimization criterion. For concreteness,
we will consider compliance
minimization, however, the framework we
present below extends straightforwardly to other forms of topology optimization problems.

Compliance minimization, invokes the
problem to
\begin{equation}
\begin{split}
    &\min_{x} C = \int_{\Gamma} dV f(u(x),x) \\
    & \text{s.t.}\ \int_{\Gamma} dV \, x = V_0 \; ,
  \label{eq:TopOpt}
\end{split}
\end{equation}
where $C$ is the compliance, $x$ is the material density over the domain, $\Gamma$, with
volume element $dV$, and total material volume $V_0$. Compliance is taken as the integral of the strain energy density, $f(u(x),x)$, 
where the displacement, $u(x)$, satisfies the stiffness equation \cite{TopOptBook}. Topology 
optimization approaches minimize compliance by making a finite-element
approximation and invoking gradient-based algorithms \cite{sigmundTopologyOptimizationApproaches2013,andreassenEfficientTopologyOptimization2011}.

\subsection{\label{Sec:PLFilter}Pareto-Laplace Filter}
We express the number of possible design solutions, $\{x\}$, with
a given compliance and fixed material volume $V$ as
$\Omega(C,V)$\footnote{Note: in the continuum, thinking of $\Omega(C,V)$ as a
  count can yield an answer that is formally infinite. In practice, $Z(\beta)$
  as defined in Eq.\ \eqref{eq:LTfamiliar} need only be defined up to an overall
  multiplicative constant, which one can use to absorb any divergence. However,
  in such cases, one could define $\Omega$ as a volume, which would amount to the
same thing.},
and apply the Pareto-Laplace filter 
\begin{equation}
  Z(\beta) = \int_{C_\text{min}(V)}^{\infty}dC e^{-\beta C} \Omega(C,V) \; ,
  \label{eq:LTfamiliar}
\end{equation}
where $\beta$ plays the role of the Laplace variable, and $C_\text{min}$ is the
minimum possible value for the compliance at fixed material volume $V$. $\Omega(C,V)$ encodes the
``volume'' of patterns of material distribution that realize a compliance of $C$ and have physical
material volume $V$. The quantity $Z(\beta)$ aggregates those volumes to give a total ``mass'' of the
solution space for all possible $C$, where the weighting factor for a
solution, like $x$, is given by $w(C) = e^{-\beta(C(x)-C_\text{min})}$.
Mathematically, $Z(\beta)$ is a generating function that is a weighted sum over
the number of ways of realizing designs of fixed compliance and material volume,
and so it encodes the geometry of the solution space \cite{ParetoLaplace}.

The interpretation of the Laplace transform in Eq.\ \eqref{eq:LTfamiliar} is
analogous to the \linebreak time/frequency domain contrasts that inform other areas of
engineering. As $\beta\to\infty$, only solutions $\{x\}$ for which
$C\approx C_\text{min}$ contribute, and hence $Z(\beta\to\infty)$ is
dominated by near minimum solutions. Hence, in the limit $\beta\to\infty$, one recovers the original
compliance minimization problem.

As $\beta\to 0$, all solutions receive equal weight, regardless of their
compliance. The rate at which $Z$ decreases as $\beta$ goes from $0$ to $\infty$
encodes the growth of $\Omega(C,V)$ with $C$, analogously to the way that
frequency space encodes transients in time domain problems.

\subsection{\label{SubSec:FilterImplementation}Filter Implementation}
Though Eq.\ \eqref{eq:LTfamiliar} geometrizes the solution space in the form of
a Laplace transform, in practice it is not possible to directly evaluate the
integral, not only because $C_\text{min}(V)$ is a primary unknown quantity in
the original formulation of the problem in Eq.\ \eqref{eq:TopOpt}, but also
because $\Omega(C,V)$ is not known. However, analogous density of states transforms
are ubiquitous in statistical physics, where
closed-form evaluation is only possible in a very small minority of cases, see e.g.\ \cite{Sethna2021}. In the majority of cases, a variety of analytic and numerical techniques facilitate working directly in terms of the underlying degrees of freedom\cite{LandauBinderMC}.

For compliance minimization, the underlying degrees of freedom are given by the
design field $\{x\}$. It is convenient to work with the discrete
approximation \footnote{Note that it is possible to express Eq.\
  \eqref{eq:LTfamiliar} via the continuum form of the design field $x$. In that
  form, $Z(\beta)$ would be a version of what is referred to as a statistical
  field theory in physics, see, e.g.\ Ref.\ \cite{KardarFields}. However, the
  field representation would require us to introduce a significant amount of
  notation that we will quickly dispense when we go to the finite element
version for ease of implementation.},
where Eq.\ \eqref{eq:LTfamiliar} can be rewritten as
\begin{equation}
    Z(\beta) = \sum_{\{x\}} e^{-\beta C} \delta\left(\sum_{e}x_e - V\right) \; .
    \label{eq:Zdef}
\end{equation}
where compliance, $C$, is evaluated for each solution ($x$), with each solution representing
a set of densities, $\{x_e\}$ and $\delta$ is a Dirac delta function that enforces the material
volume constraint.

Eq.\ \eqref{eq:LTfamiliar}, and the equivalent form Eq.\ \eqref{eq:Zdef} geometrize the compliance-minimization solution
space. In statistical physics $\beta=1/T$, where $T$ is thermodynamic temperature. There are
well-defined computational techniques, see, e.g.\ Ref.\ \cite{frenkelsmit},
that provide statistical sampling that can generate the distributions of $\{x\}$
that contribute to $Z(\beta)$ up to arbitrary accuracy.

\subsection{\label{SubSec:PhysicalStatisticalInterpretation}Physical and Statistical Interpretation}
Laplace transforms, Eq.\ \eqref{eq:Zdef}, are typically not possible to
evaluate in closed form. However, as a mathematical construction, Eq.\
\eqref{eq:Zdef} has the form of a partition function used for more than a
century in statistical physics \cite{Sethna2021}. Notably, along with the thermodynamic interpretation of $Z(\beta)$, the connection between thermodynamics and statistical mechanics, see, e.g.\ Refs.\
\cite{LLv5, Sethna2021}, dictates that $Z(\beta)$ can also be interpreted as a
generating function for probability distributions. From a probabilistic point of
view, that facilitates the computation of statistical averages over the solution
space, e.g.,
\begin{equation}
  \left<C\right> = \frac{1}{Z(\beta)}
  \int_{C_\text{min}}^\infty dC e^{-\beta C} \Omega(C,V) C \; ,
  \label{eq:Cavg}
\end{equation}
where the notation $\left< \cdot \right>$ indicates the so-called thermal or
ensemble average. Note, that by inspection we must have that
\begin{equation}
  \left<C\right> = -\frac{\partial \ln(Z(\beta))}{\partial \beta} \; ,
  \label{eq:CdZ}
\end{equation}
which provides a clearer indication of how $Z(\beta)$ generates
a probability distribution \cite{ParetoLaplace}.

\subsection{\label{SubSec:NOE}Near Optimal Expansion}

Considering the global minimum at some 
compliance $C_\text{min}$, then the volume of the solution space will vanish for
$C<C_\text{min}$. In that case, suppose that for values of compliance that
fall just above $C_\text{min}$ we can approximate the solution space volume
\begin{equation}
  \Omega(C,V) \propto (C-C_\text{min})^{\frac{N_{IP}}{\nu}-1} \; ,
  \label{eq:OmegaCApprox}
\end{equation}
where $N_{IP}$ is the effective number of degrees of freedom in play near the minimum
and $\nu$ is an index of growth. In this case, we can evaluate Eq.\
\eqref{eq:LTfamiliar} directly, which gives us
\begin{equation}
  Z(\beta) \propto e^{-\beta C_\text{min}} \beta^{\frac{N_{IP}}{\nu}} \; ,
  \label{eq:Zapprox}
\end{equation}
where we have dropped negligible overall multiplicative constants. That means
that in the limit of large $\beta$ or, equivalently, low temperature $T$, we can
estimate Eq.\ \eqref{eq:CdZ} to be
\begin{equation}
  \left< C\right> \approx C_\text{min} + \left(\frac{N_{IP}}{\nu}\right) T \; .
  \label{eq:ClowT}
\end{equation}

\subsection{\label{SubSec:BNOD}Beyond Near-Optimal Design}
Consider the case where some degrees of freedom are ``in play'' near the
minimum, up to some saturation point, $C_*$, after which a
different set of degrees of freedom comes into play. We represent this by
\begin{equation}
  \Omega(C,V) =
  \begin{cases}
    \gamma_< (C-C_\text{min})^{\frac{N_<}{\nu}-1} & C<C_* \\
    \gamma_< (C_*-C_\text{min})^{\frac{N_<}{\nu}-1}+
    \gamma_> (C-C_*)^{\frac{N_>}{\nu}-1} & C>C_* \\
  \end{cases}
  \; ,
  \label{eq:Opiece}
\end{equation}
where $\gamma_{<,>}$ are geometric coefficients, and $N_{<,>}$ are scaling
exponents giving the number of effective degrees of freedom for $\Omega(C,V)$
on either side of $C_*$.

Integrating Eq.\ \eqref{eq:Opiece} gives
\begin{equation}
  \begin{split}
    Z(\beta) =&
    e^{-\beta C_\text{min}}\beta^{\frac{N_<}{\nu}}
    \gamma_<
    \left[\Gamma\left(\frac{N_<}{\nu}\right)-\Gamma\left(\frac{N_<}{\nu},\beta(C_*-C_\text{min})\right)\right]
      +\\
      & e^{-\beta C_*}\beta^{\frac{N_>}{\nu}}
      \left[
        \gamma_>\Gamma(\frac{N_>}{\nu})
        +\gamma_<\left(\beta(C_*-C_\text{min})\right)^{\frac{N_>}{\nu}}
      \right] \; .
    \end{split}
  \label{eq:Zpiece}
\end{equation}
where $\Gamma(\cdot,\cdot)$ is the incomplete Gamma function.

We can use this to find that for
$\beta(C_*-C_\text{min})\gg 1$
(equivalently $T\ll C_*-C_\text{min}$),
\begin{equation}
  \left< C\right>\approx
  C_\text{min}+ \frac{N_<}{\nu\beta}
  = 
  C_\text{min}+\left(\frac{N_<}{\nu}\right)T
  \; ,
  \label{eq:OTsPiece}
\end{equation}
whereas for
$\beta(C_*-C_\text{min})\ll 1$
(equivalently $T\gg C_*-C_\text{min}$),
\begin{equation}
  \left< C\right>\approx
  C_* + \frac{N_>}{\nu\beta}
  =
  C_*+\left(\frac{N_>}{\nu}\right)T
  \; .
  \label{eq:OTlPiece}
\end{equation}

Note that for $T\ll C_*-C_\text{min}$ and
$T\gg C_*-C_\text{min}$,
$C$ asymptotes to linear response in $T$. In both cases,
the slope is determined by the power law growth of $\Omega(C,V)$. The
exponent in this power law growth is, in turn, determined by the number of
degrees of freedom that are ``in play'' at that level of $T$ (or $\beta$).

\subsection{\label{SubSec:NumericalIntegralImplementation}Numerical Integral Transform Implementation}
Casting the Pareto-Laplace transform of Compliance minimization in the
equivalent form of Eq.\ \eqref{eq:Zdef} unlocks a vast array of techniques in
physics, chemistry, and materials science that developed to handle analogous
problems. 

To generate the material distributions $\{x_e\}$ that contribute to Eq.\
\eqref{eq:Zdef}, we identify $Z(\beta)$ as a partition function with compliance
as the potential energy $U = C(\{x\})$, and fixed temperature
$T=\frac{1}{\beta}$. We treat material densities as ``particles'', where each
particle is fixed to a position $x_e$ in the design domain, and its ``position''
in an auxiliary dimension describes the density of material in that cell (no
material : $x_e=0$, filled cell: $x_e=1$). It will be convenient to visualize
the ``filling'' dimension as perpendicular to the design domain. 

With this framing, the task of computing $\Omega(C,V)$ is
equivalent to generating a volume-preserving series of solutions.
However, since we are interested instead in computing $Z(\beta)$, we must
generate a series of configurations appropriate to that quantity. For systems of conventional particles, this task represents the central
challenge in molecular simulation, and it is one of the first problems that was
tackled with scientific computing, so a vast literature has emerged over the
last several decades \cite{frenkelsmit,ReversibleNHT}. As such, a variety of
means exist for computing Eq.\ \eqref{eq:Zdef}, and so the approach that we
develop here is a powerful but first means of obtaining traction.

Our approach implements the computation of Eq.\ \eqref{eq:Zdef} by exploiting an
analogy with industry-standard implementations of the framework of molecular
dynamics used in chemistry, physics, and materials science. The approach we take
here for topology optimization most closely mirrors the treatment that Ref.\
\cite{ReversibleNHT} develops for conventional systems. For more details on the numerical implementation of the integral transform using molecular
dynamics for topology optimization, refer to Appendix \ref{AppSec:NumericalDetails}.

\begin{acknowledgments}
We thank I.\ Babayan for useful discussions and comments on the manuscript, and
R.\ Perez for discussions that initiated this investigation and collaboration at an early stage of the project. We
acknowledge the support of the Natural Sciences and Engineering Research Council
of Canada (NSERC) grants RGPIN-2019-05655 and DGECR-2019-00469. Computations
were performed on resources and with support provided by the Centre for Advanced
Computing (CAC) at Queen's University in Kingston, Ontario. The CAC is funded
by: the Canada Foundation for Innovation, the Government of Ontario, and Queen's
University.
\end{acknowledgments}
\appendix 
\renewcommand{\theequation}{A\arabic{equation}}
\setcounter{equation}{0}
\renewcommand{\thefigure}{A\arabic{figure}}
\setcounter{figure}{0}

\section{\label{AppSec:NumericalDetails}More Details on Numerical Integral Transform Implementation} 
A full description of molecular dynamics is beyond the scope of this manuscript,
but the key to computing Eq.\ \eqref{eq:Zdef} is the introduction of two sets of
auxiliary variables. For defining the first set of auxiliary variables, if we regard the density of material
at each point as the ``position'' of a particle that determines potential energy
(vis-a-vis compliance), then it is possible to introduce a kinetic energy for the
particles that associates a momentum with each position. One can then multiply
Eq.\ \eqref{eq:Zdef} by one in the form of
\begin{equation}
  1 =
  \frac{\prod_{x_e} \int_{-\infty}^{\infty}dp_e e^{-\beta p_e^2}}
  {Z_\text{free}(\beta)} \; ,
  \label{eq:Zfree}
\end{equation}
where the factor in the numerator corresponds with the definition of
$Z_\text{free}$ in the denominator. It is straightforward to show that
$Z_\text{free}=(\tfrac{\pi}{\beta})^{N/2}$, where $N$ is the number of elements
to deposit material in. With this factor, Eq.\ \eqref{eq:Zdef} is equivalent to
\begin{equation}
    Z(\beta) =
    \left(\frac{\beta}{\pi}\right)^{\frac{N}{2}}
    \sum_{\{x,p\}} e^{-\beta(\sum_e p_e^2+C)}
    \delta\left(\sum_{e}x_e - V\right) \; ,
  \label{eq:Zkin}
\end{equation}
where the sum should be interpreted schematically. Note that by adding degrees
of freedom that have the form of momentum, Eq.\ \eqref{eq:Zkin} is
mathematically equivalent to our original quantity of interest, however, by
explicitly introducing momentum we are able to induce dynamics to generate
configurations of the design field.

Ensuring that the induced dynamics generates configurations of the design field
that correctly reproduce the integral in Eq.\ \eqref{eq:LTfamiliar} requires
introducing a second set of auxiliary variables. The integrand in Eq.\
\eqref{eq:Zkin} depends on the quantity
\begin{equation}
  \mathcal{H} \equiv \sum_e p_e^2 + C
  \label{eq:Ham}
\end{equation}
which has the form of a Hamiltonian in classical mechanics, e.g.\ c.f.\ Ref.\
\cite{LLv1}. Hamiltonians of this form generate dynamics that conserve energy,
however the form of Eq.\ \eqref{eq:Zdef} indicates that we require instead
dynamics that enact constant temperature $T=1/\beta$.

There are several methods, see, e.g.\ Ref.\ \cite{frenkelsmit} to enact
constant $T$. The one that we will adopt here, a so-called Nos\'e-Hoover chain
thermostat, is widely used because it performs well in a wide variety of systems
and it scales to very large systems. Scaling considerations are important
because they signal the prospect for high performance implementations of the
current approach to the giga-voxel resolution that has been realized with
conventional topology optimization \cite{gigavoxel}.

The Hamiltonian of the system $\mathcal{H}$ with one constraint (the volume
constraint) which is coupled with a Nos\'{e}-Hoover chain of length $N_c$ is
\begin{equation}
\label{eq::NoseHooverHamiltonian}
\mathcal{H}_{NHC} = \mathcal{H} + \sum_{k=1}^{N_c} \frac{1}{2}Q_k v_{s_k}^2+(N-1)T_{target}s_1+\sum_{k=2}^{N_c}T_{target}s_k
\end{equation}
where k\textsuperscript{th} Nos\'{e} position, $s_k$, is defined as,
\[
\dot s_k = v_{s_k}; \quad k=1,\ldots,N_x
\]
where $v_{s_k}$ is the k\textsuperscript{th} Nos\'{e} velocity, $Q_k$ is the
k\textsuperscript{th} Nos\'{e} mass, $T_{target}$ is the temperature we want to
hold the system in, and it is understood that $v_{s_{M+1}}=0$
\cite{ReversibleNHT, Tuckerman1992, GlennMartyna1996}. Working from Eq.
(\ref{eq::NoseHooverHamiltonian}), Hamilton's equations give the equations of
motion for all particles as first-order differential
equations\cite{ReversibleNHT, LLv1}.

In general terms, the Nos\'e-Hoover chain thermostat works by coupling the
``particles'' that represent the material distribution to an auxiliary set of
``particles'' that function as a heat bath. The material distribution particles
interact with the heat bath particles in a specific manner that allows for an
exchange of energy between the two that replicates the effectiveness of constant
temperature.

A stable numerical implementation of integrating the equations of motion can be
achieved using a symplectic integrator \cite{ReversibleNHT,Yamamoto2006} and the
Lagrange multiplier must be computed at each time-step such that the particle
positions (densities) satisfy the volume constraint \cite{ReversibleNHT}. 
\newpage

\section{\label{AppSec:SitesDynamics}Sites Dynamics}
\begin{figure}[!h]
\centering
  \includegraphics[width=1.0\textwidth]{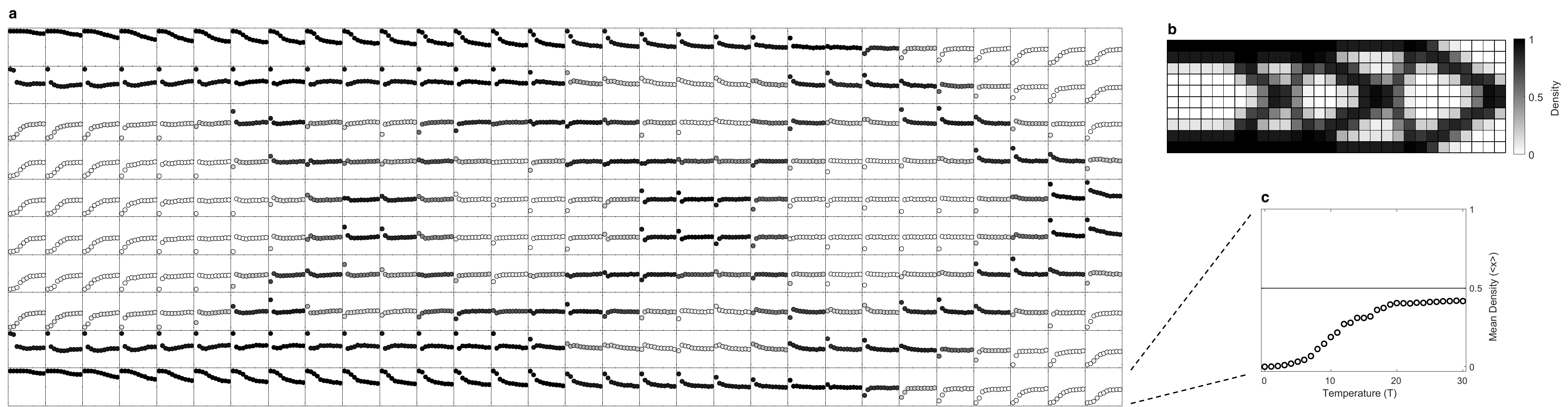}
  \caption{
    \textbf{Mean density vs $T$ for all sites for a cantilever beam in 2D.} Fig.\ \ref{Fig:CondensationPhases} (main
    text) panels f,h,j,l showed mean density vs $T$ for selected material sites.
    (\textbf{a}) plots mean density vs $T$ for all material sites of the
    cantilever beam design which is shown in (\textbf{b}). Data points in (\textbf{a}) are shaded
    by the density of the corresponding site in the ultimate design in (\textbf{b}). (\textbf{c})
     magnifies the results from a single site to show the density and
    temperature scales used throughout the individual site plots in (\textbf{a}).
  }
  \label{Fig:DensityDynamics}
\end{figure}

\begin{figure}[!h]
\centering
  \includegraphics[width=1.0\textwidth]{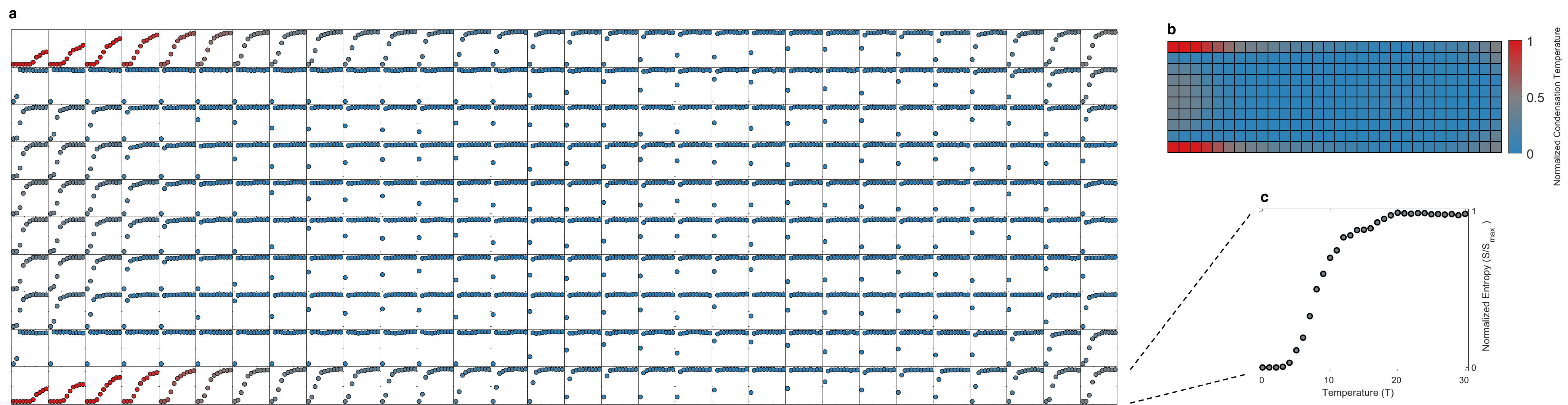}
  \caption{
    \textbf{Normalized entropy density vs $T$ for all sites for a cantilever beam in 2D.}
    Fig.\ \ref{Fig:CondensationPhases} (main text) panels g,i,k,m showed entropy density vs $T$ for
    selected material sites. (\textbf{a}) plots entropy density vs $T$ for all
    material sites of the cantilever beam design shown in (\textbf{b}). Data points
    in (\textbf{a}) are shaded by the density of the corresponding site in the
    ultimate design in (\textbf{b}). (\textbf{c}) magnifies the results from a single site to
    show the entropy and temperature scales used throughout the individual site
    plots in (\textbf{a}).
  }
  \label{Fig:EntropyDynamics}
\end{figure}
\newpage
\section{\label{AppSec:EntropyCalculation}Entropy Calculation}
The entropy of a site's density, $x_e$, is calculated using the probability distribution of the site's density within the ensemble of solutions at a given temperature. In the continuous limit, the entropy is expressed as
\begin{equation} S(x_e) = - \int_0^1 P(x')\log(P(x'))dx' ; , \end{equation}
where $P(x)$ represents the probability distribution of the site's density, $x_e$, and $x'$ is the integration variable.

\begin{figure}[!h]
\centering
  \includegraphics[width=1.0\textwidth]{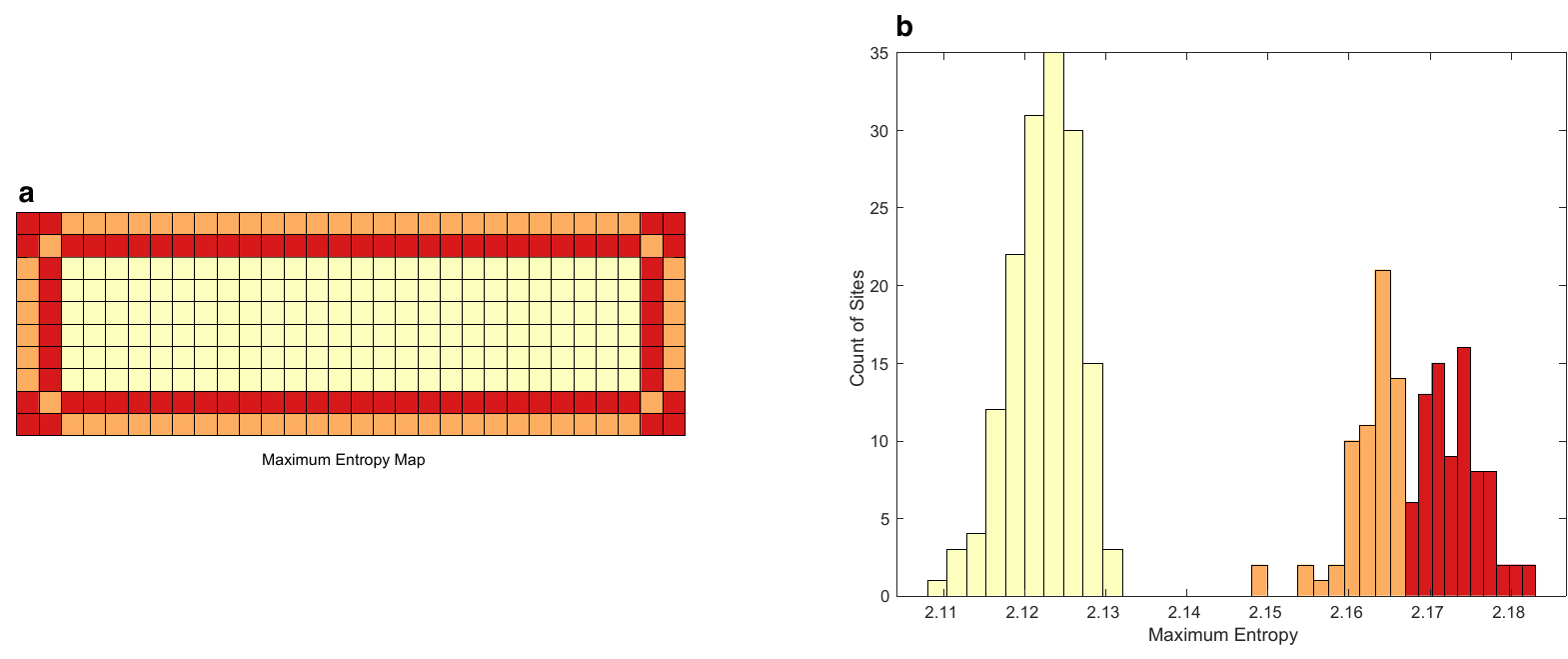}
  \caption{
    \textbf{Distribution of maximum entropy by site.} Site-specific entropy density is a
    useful indicator for the condensation process that drives morphogenesis.
    However, a careful analysis of entropy density at high temperature shows
    that there is an interaction between the filter and the boundary
    conditions on the region that lead to a difference in entropy density maxima
    that varies by $\lesssim 5\%$ across sites. This variation is plotted in
    (\textbf{a}), and aggregated as a histogram in (\textbf{b}).
  }
  \label{Fig:EntropyCalculation}
\end{figure}

At high temperatures, the entropy reaches its maximum value, corresponding to the absence of a gradient force from the objective function. This condition can be achieved by setting the gradient of the objective function to zero. The maximum entropy values for the cantilever beam problem, as studied in this paper, are shown in Fig.\ \ref{Fig:EntropyCalculation}.

The differences observed in the maximum entropy values stem from the use of a density filter, a common practice in topology optimization to mitigate numerical instabilities. 
\newpage
\section{\label{AppSec:3DCantileverBeam}3D Cantilever Beam}
\begin{figure}[!h]
\centering
  \includegraphics[width=1.0\textwidth]{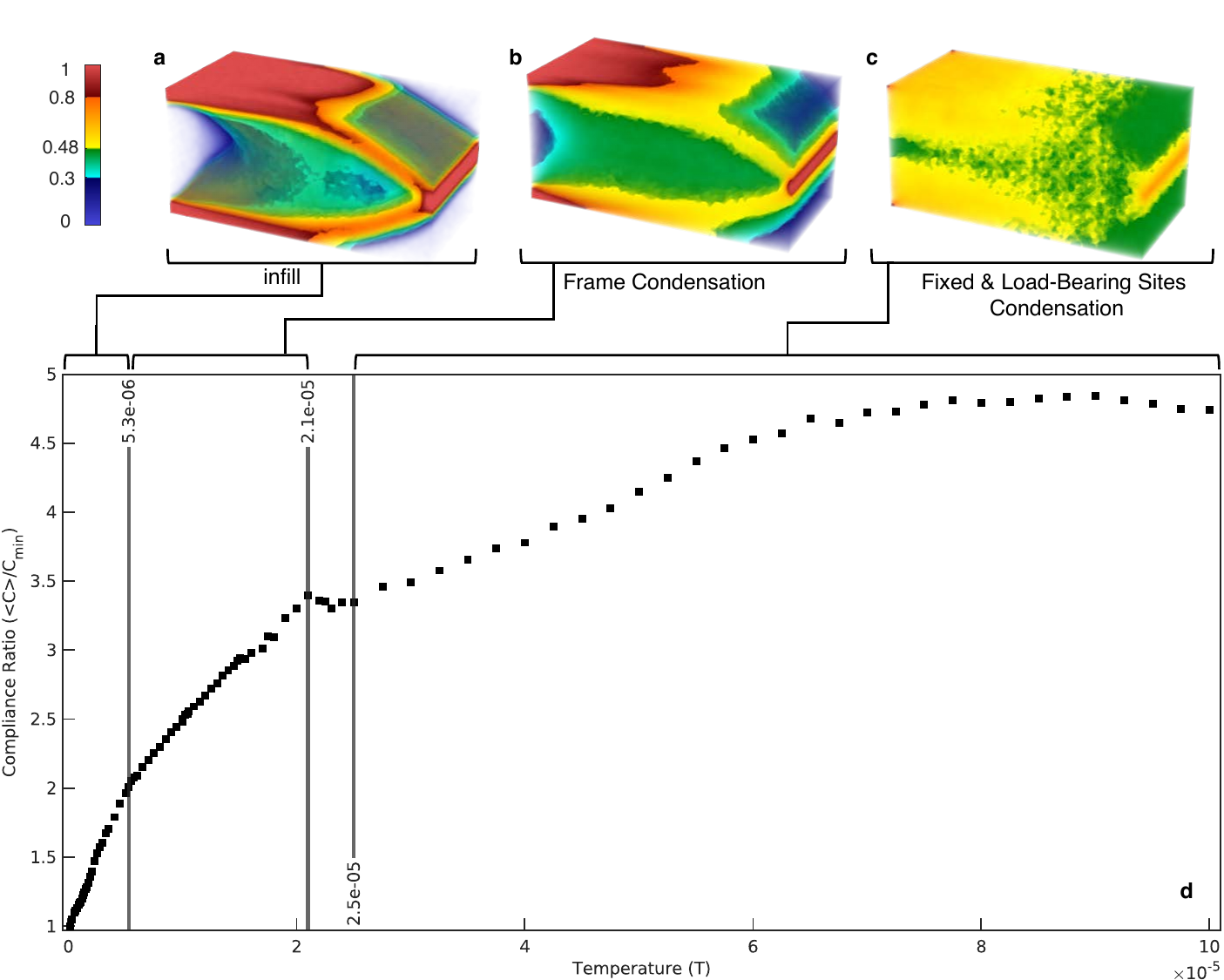}
  \caption{
    \textbf{Condensation process for morphogenesis of a 3D cantilever beam.} (\textbf{a-c})
    show material density using a multicolour material density scale that
    has been adjusted to maximize illustration of structural details. (A
    conventional colour scale analogue is shown in the main text.) (\textbf{d}) shows
    $\left<C\right>$ vs $T$, which indicates a similar set of condensation
    regimes as the 2D cantilever beam, albeit at lower $T$.
  }
  \label{Fig:Phases3D}
\end{figure}

\begin{figure}[!h]
\centering
  \includegraphics[width=1.0\textwidth]{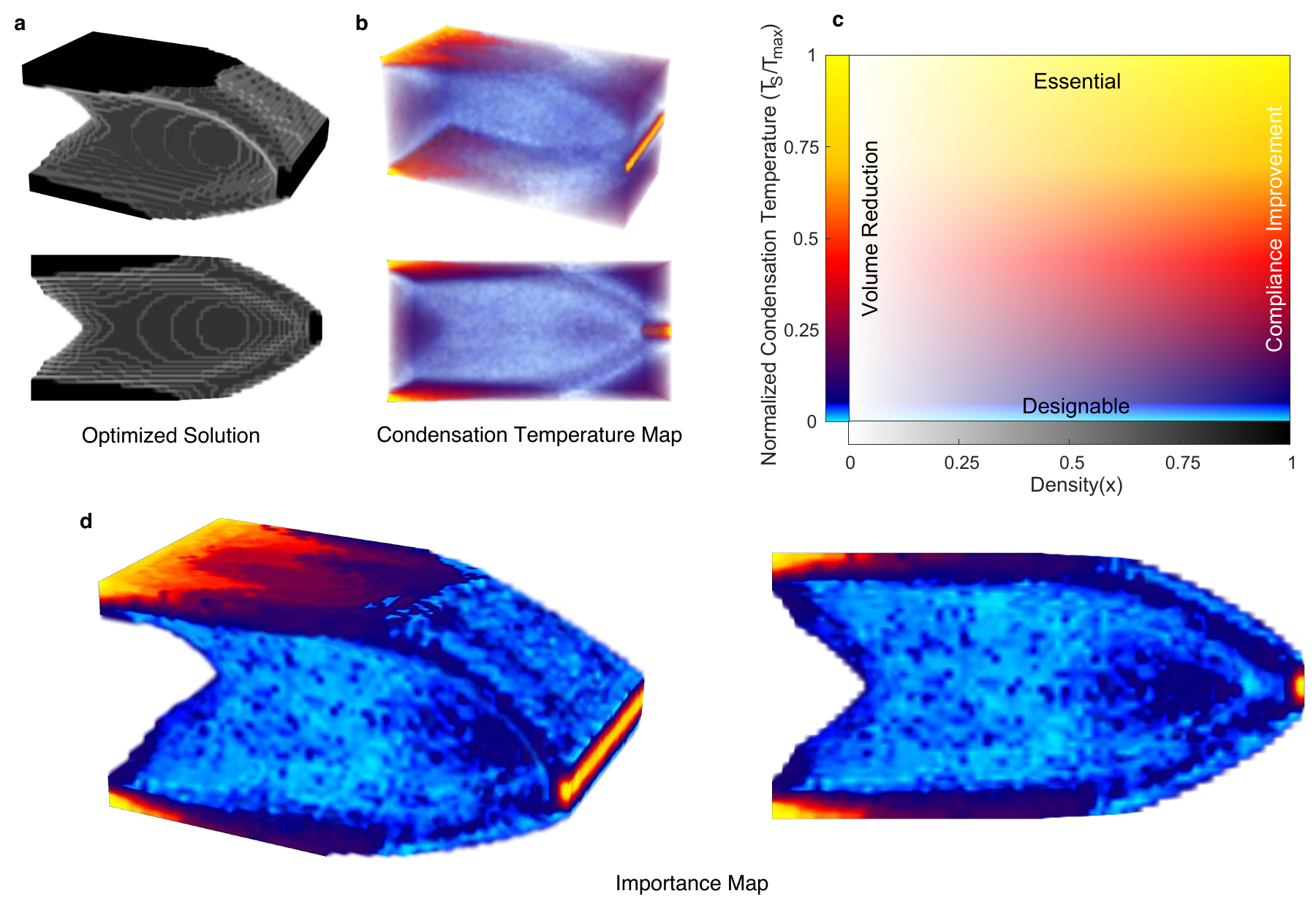}
  \caption{
    \textbf{Optimized solution, condensation temperature map, and importance map for 3D cantilever beam
    solution.} (\textbf{a, b, d}) illustrate alternate views in addition to those
    shown in the main text, which are shaded using a colourmap (\textbf{c}) to facilitate
    identification of structure.
  }
  \label{Fig:ImportanceMap3D}
\end{figure}

%

\end{document}